\documentstyle[epsfig,11pt]{article}
\begin{document}
\begin{flushright}SJSU/TP-98-19\\November 1998\end{flushright}
\vspace{1.7in}
\begin{center}\Large{\bf Space-time counterfactuals}\\
\vspace{1cm}
\normalsize\ J. Finkelstein\footnote[1]{
        Participating Guest, Lawrence Berkeley National Laboratory\\
        \hspace*{\parindent}\hspace*{1em}
        e-mail: JLFINKELSTEIN@lbl.gov}\\
        Department of Physics\\
        San Jos\'{e} State University\\San Jos\'{e}, CA 95192, U.S.A
\end{center}
\begin{abstract}
A definition is proposed to give precise meaning to the counterfactual
statements that often appear in discussions of the implications of
quantum mechanics.  Of particular interest are counterfactual statements
which involve events occurring at space-like separated points, which do
not have an absolute time ordering.  Some consequences of this
definition are discussed.

\end{abstract}
\newpage
\section{Introduction}
Discussions of the interpretation or the implications of quantum mechanics
often use the language of counterfactuals (see for example refs. [1]--[14]); 
however, there does not seem
to be agreement on exactly what these counterfactuals are to be taken
to mean.  In this paper I will propose a precise definition for  
counterfactual statements which are used in discussions of quantum theory.

Consider this simple situation: I have two particles called $A$ and $B$,
each of spin $\frac{1}{2}$,
initially prepared in a state of total spin zero.
First I measure $S_x$ (the x-component of the spin) of $A$, and find
the value +1 (expressed in units of $\frac{1}{2}\hbar $);
then I measure the value of $S_y$ of $B$, and also find the value +1.
I suspect that most of us would consider, in this situation, the
statement ``If I had measured $S_x$, instead of $S_y$, for $B$, I would
have obtained the value -1'' to be true, and likewise the statement
``If I had measured $S_x$, instead of $S_y$, for $B$, I would
have obtained the value +1'' to be false. But what do these statements,
about a situation that does not exist, really mean?  By the antecedent
``If I had measured $S_x$, instead of $S_y$, for $B$'' I am imagining
a situation which is exactly like the actual one up until the time
of the measurement of the spin of $B$ (in particular, a situation in
which the particles did indeed have initial total spin zero,  
and in which I did indeed first measure $S_x$ for $A$ and obtained the
value +1), and in which the value of $S_x$ of $B$ is then measured.
The correct conclusion ``I would have obtained the value -1''
can be understood as simply being the implication of quantum theory 
for this imagined situation.

In the next section I will propose a definition of counterfactual
which will formalize this understanding of this simple example,
and which will permit an extension to a class of counterfactuals
which I call space-time counterfactuals.  This class will include
statements about situations more complicated than the one discussed 
above, for example situations in which several choices are made; 
most importantly, 
it will include counterfactual statements about
events at different locations, for which a Lorentz-invariant notion
of temporal precedence does not exist.  I believe that most of the
counterfactuals used in discussions of quantum theory fall into this
class, but of course many counterfactuals which are used in other 
contexts do not.  I am certainly not  presenting a new general theory
of counterfactuals; the definition I will present is essentially
the application of the theory of counterfactuals of Lewis \cite{L}
to space-time counterfactuals,
but with a modification of that theory to accommodate the lack of a
relativistically-invariant time ordering.

I will restrict the discussion to counterfactuals which concern localized
macroscopic phenomena; I will
refer to such phenomena generically as ``events'', and will employ
the idealization that an event occupies a single point of space-time.
Of course I will allow statements about combinations of events; the
restriction is that each event has a definite location.  In the simple
example given above, the event in the antecedent was the choice of which
measurement to make; the class of counterfactuals I consider also includes
those whose antecedent event is, for example, the {\em result} of a
measurement performed on a quantum system.  I will
certainly want to discuss situations in which particles at different
locations are in a quantum-entangled state, which therefore cannot
be represented by any combination of properties each of which refers
only to a single location.  Even in these situations, I choose to 
consider only those statements which refer to macroscopic, and hence
localizable, properties.  Strictly speaking, the simple example
discussed above, in which I specified the quantum state of the 
two-particle system, does not respect this restriction. However, I can
consider the specification of the quantum state as merely an abbreviation
for a description of the macroscopic procedure by which it was
prepared.  Saying this does not imply a commitment to any interpretation
(e.\ g., an instrumentalist interpretation) of quantum theory; it is 
merely a restriction on what I choose, for the present purposes, to
talk about.

I will be using, starting in the next section, a language involving the
assessment of the similarity of a ``possible world'' to the actual world.
As a generalization of the statement in the  simple example that
``everything is the same up to the time of the measurement on $B$'',
I will base the assessment on the region of space-time over which the
two compared worlds agree,  without introducing any notion of
a ``degree'' of disagreement at individual space-time points.
Of course such a  notion could be important for other applications of
counterfactual statements, but I limit the discussion here to
cases in which it is not.  So the ``space-time counterfactuals'' 
I consider are those
which involve  only localizable (in fact, macroscopic) properties,
and whose truth can be assessed by considering the regions of space-time
in which those properties are, or are not, obeyed.

In the next section I will define counterfactuals in a way which is 
appropriate for this case.  In the final section I will present some
applications of this definition, and point out some problems with some
alternative definitions.  

\section{Definition of counterfactual statements}
\def\cf{\raisebox{3 pt}{\fbox{}}\!\! \rightarrow}
I want to adapt the analysis of Lewis \cite{L} to space-time
counterfactuals.  That analysis uses the semantics of ``possible worlds.''
We can a take possible world to be any  which is consistent with the laws
of physics (in practice, it is usually the laws of quantum physics with
which we will be concerned); no commitment to the existence of any world
other than the one actual world is required. 

Lewis' analysis requires a judgment as to the comparative similarity
of various possible worlds to the actual world.  Call a world in
which a proposition $\phi$ is true a ``$\phi$-world''. Then, in a 
sense which will be made precise below, the counterfactual which is
written $\phi \; \cf \psi$, and which is read ``If $\phi$ were true,
then $\psi$ would be true'' is taken to mean that $\psi$ is true in
the $\phi$-worlds most similar to the actual world; it is not required
that $\psi$ be true in {\em all} possible $\phi$-worlds. In the simple
example discussed in the previous section, $\phi$ was ``$S_x$ of $B$
was measured'' and $\psi$ was ``Value -1 was obtained''.
Letting $t$ be the time of the measurement of  $B$, and not yet worrying
about the relativistic absence of an unambiguous time ordering, we said
in effect that $\phi \; \cf \psi$ was true because $\psi$ was true in
all $\phi$-worlds which agreed with the actual world at all times
earlier than $t$.  To express this in Lewis' language, we would say
that all worlds which agree with the actual world at all times earlier
than $t$ are equally similar to the actual world, and are more similar to
it than is any world which differs from the actual world at any time
earlier than $t$.   

To see how to generalize this judgment to the relativistic case,
consider three space-time points called $A$, $B$, and $C$, such that
$B$ is unambiguously later than $A$, and that $C$ is space-like
separated from both $A$ and $B$.  This situation is depicted, using
the coordinates of some particular Lorentz frame, in figure 1, which also
indicates the forward light-cones of these three points.  
\begin{figure}
\epsffile{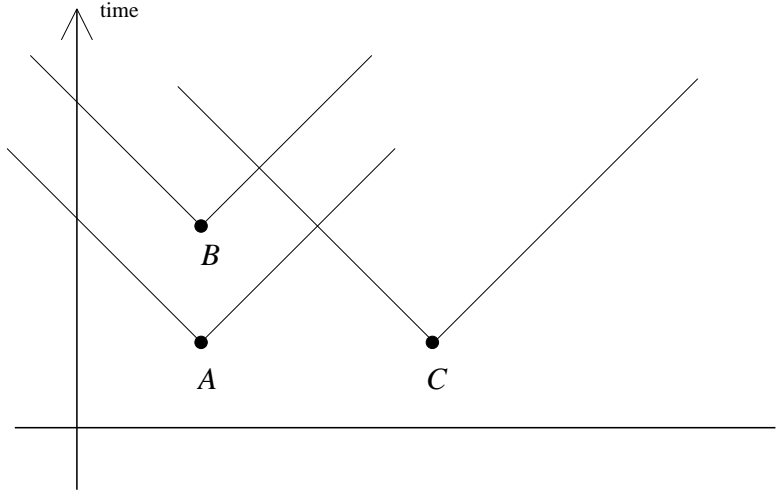}
\caption{Three space time points $A$, $B$, and $C$, and their forward
light cones. $B$ is in the 
unambiguous future of $A$, and  $C$ is space-like separated from both
$A$ and $B$.}
\end{figure}
Consider also
three possible worlds: world $W_A$ first differs from the actual
world at point $A$; world $W_B$ first differs from the actual world
at $B$; likewise world $W_C$ first differs at $C$.  From the preceding
discussion, we would say that $W_B$ is more similar to the actual world than
is $W_A$, since it first differs from the actual world at an (unambiguously)
later time.  Comparing $W_A$ with $W_C$, there is no way to judge
between them; in fact in the frame shown they each differ from the
actual world at the same time. We might therefore be tempted to judge
that $W_A$ and $W_C$ are equally similar to the actual world.  But wait!
Since $C$ and $B$ are space-like separated, there is a frame in which
$W_C$ and $W_B$ first differ from the actual world at the same time,
so we should be just as tempted to judge that $W_C$ and $W_B$ are
equally similar to the actual world.  However, if we judge that
$W_A$ and $W_C$ are equally similar, and also that $W_C$ and $W_B$
are equally similar, we would be forced to conclude that $W_A$ and
$W_B$ are equally similar, in conflict with the judgment that $W_B$
is {\em more} similar than is $W_A$.

What to do? We could arbitrarily pick a Lorentz frame, and declare
that the possible world which, in that frame, deviates from the
actual world at the latest time is the most similar; we could then
use this judgment of similarity to define counterfactuals as in the
analysis of Lewis. However, this definition will lead to the situation
that the truth of certain counterfactual statements will depend on
which frame we happened to pick (an example of this will be presented
in the next section).  It is not inconsistent to say that the truth
of a statement may depend on the choice of a frame (the statement
``the earth is at rest'' is true in some frames and not in others),
but we should be able to do better. Suppose we say that a
counterfactual statement 
$\phi \; \cf \psi$ is ``verified'' in a given frame if 
$\psi$ is  true in the $\phi$-world which deviates at the latest time
in {\em that} frame, and then say that the statement is true if
there is {\em any} frame in which it is verified.  This would
certainly give us a frame-independent criterion for truth of a counterfactual,
but at the price of, for example, permitting situations
in which two counterfactuals $\phi \; \cf \psi_1$ and
$\phi \; \cf \psi_2$, with $\psi_1$ and $\psi_2$ contradictory,
would both be true. We will see an example of this in the next section. 

Without further ado, here is my proposal for a definition of 
space-time counterfactual statements.  For any space-time point $r$,
let $F(r)$ be the set consisting of $r$ itself and of all points
which are unambiguously at a later time than $r$ (that is, points
on or within the forward light-cone of $r$). For any set of points
$D$, let $\overline{D}$ be the union of $F(r)$ over all $D$; that is,
\[\overline{D}:=\{ r'|r'\in F(r) \mbox{ for some }r \in D \}. \]
Call $\overline{D}$ the ``future closure'' of $D$, and note that
for any $D$, $\overline{\overline{D}}=\overline{D}$.
Now let $W_a$ denote the actual world, and $W_p$ denote some other possible
world. Let $D_p$ be the set of space-time points at which $W_p$ differs
from $W_a$, that is, the set of points at which there is an event which
is different in $W_p$ than in $W_a$. The definition of a space-time
counterfactual which I propose, and which I will refer to as DSTC, is:

\vspace{2mm}
The statement  $\phi \; \cf \psi$
is true when either
\begin{enumerate}
  \item There are no possible $\phi$-worlds, or
  \item For any $\phi$-world $W_q$ in which $\psi$ is not true,
    there is a $\phi$-world $W_p$ in which $\psi$ {\em is} true,
    with $\overline{D_p}\subset \overline{D_q}$.  
\end{enumerate}

In the second condition above, I am taking 
$\overline{D_p}\subset \overline{D_q}$ to require that 
$\overline{D_p}$ be a {\em proper} subset of  $\overline{D_q}$.
This definition DSTC indicates that I regard  $\overline{D_p}$ 
as a measure  of the similarity of $W_p$ to $W_a$; the condition
$\overline{D_p}\subset \overline{D_q}$ could be read as
``$W_p$ is more similar to $W_a$ than is $W_q$.'' 
In fact, DSTC
closely follows the definition given by Lewis \cite{L}, except that
the use of set inclusion here means that we have a {\em partial}
(pre-)order on the set of possible worlds (not every pair of possible
worlds can be compared), while Lewis requires that the set of
possible worlds have a  {\em total} 
(pre-)order\footnote[1] 
{An alternative to DSTC, 
``$\phi \; \cf \psi$ when there exists a $\phi$-world $W_p$ in which $\psi$
is true such that, for every   $\phi$-world $W_q$ in which $\psi$ is
{\em not} true, $\overline{D_p}\subset \overline{D_q}$,'' which follows
even more closely the formulation given by Lewis, does not work with
a partial order.  Also, it should be understood that, although it is
not indicated here explicitly, the truth of a counterfactual depends
on the actual world from which it is issued (as is stated by Lewis).}.
We have a partial order because we want
similarity to
reflect the earliest time at which a possible
world differs from the actual world, and in a relativistic context
temporal precedence is a partial (rather than a total) order on
the set of space-time points. The existence of a partial order for
possible worlds has also been suggested by Pollock\cite{JLP},
on very different grounds.

There is a (pathological) situation in which DSTC would
yield an unreasonable answer: suppose that the possible $\phi$-worlds
could be arranged in an infinite sequence, say $W_i$ for
$i=1,2,3,...$ with $\overline{D_{i+1}}\subset \overline{D_i}$,
and with $\psi$ true in $W_i$ iff $i$ is even.  Then both
$\phi \; \cf \psi$ and $\phi \; \cf (\sim\psi)$ would be true.
This situation obviously would not arise if there were only a
finite number of possible worlds. Since in essentially all applications
the number of worlds considered is in fact finite (and often is quite
small) we would not lose much generality if we were to restrict
ourselves to cases of finite numbers of worlds. However, an even weaker
condition will suffice.  For a possible $\phi$-world $W_p$, say
that $W_p$ is a ``primary'' $\phi$-world if there is no possible
$\phi$-world $W_q$ with $\overline{D_q}\subset \overline{D_p}$.
Now say that a proposition $\phi$ is ``closed'' if,
for every possible $\phi$-world $W_q$,
either $W_q$ is itself primary, or else
there is a primary $\phi$-world $W_p$  
with $\overline{D_p}\subset \overline{D_q}$. Then DSTC should be 
understood as 
applying only to antecedents $\phi$ which are closed.  
This restriction is related to the Limit Assumption discussed
(but neither needed nor adopted) by Lewis. Essentially any space-time
proposition
$\phi$ which we might reasonably want to consider will turn out to be
closed, so that DSTC will apply; in the next section we will see
examples of propositions which can be proven to be closed.  

The definition DSTC can now be
restated as follows: For a closed proposition $\phi$,
the statement ``$\phi \; \cf \psi$'' means that either
1) there are no possible $\phi$-worlds, or 2) $\psi$ is true in
every primary $\phi$-world. With this definition, 
some but not all of the usual inferences for conditionals are valid.
Some examples are: $(\phi \; \cf \psi_a)$ and 
$(\phi \; \cf \psi_b)$ implies
$\phi \; \cf (\psi_{a}\wedge \psi_{b}) $;
$\phi \; \cf \psi_a$ implies  $\phi \; \cf (\psi_{a}\vee \psi_{b})$;
but $\phi_{a} \; \cf \psi$ does not imply
$(\phi_{a}\wedge \phi_{b}) \; \cf \psi$ (as it also does not in
Lewis' theory).  Let me record here a proof of the first example:
Assume that $\phi \; \cf \psi_a$
and that $\phi \; \cf \psi_b$.  
This means that both $\psi_a$ and $\psi_b$ are true in every
primary $\phi$-world; hence $\psi_{a}\wedge \psi_b$ is true in
every primary $\phi$-world, and hence the statement
$\phi \; \cf (\psi_{a}\wedge \psi_b)$ is true.

Since $D_p$ is the set of points at which $W_p$ differs from $W_a$,
it might be thought that $D_p$, rather than $\overline{D_p}$, should
be taken as the measure of similarity between $W_p$ and $W_a$.
But even in the simple, non-relativistic example discussed in the
first section, for which (I believe) the interpretation of the
counterfactual was uncontroversial, that is not what happened.
I expect that my reader agreed with me that what the counterfactual
meant in that example
was that the conclusion should hold in any situation
(i.\ e.\ any world) which is exactly like the actual situation up until
the time of the measurement of $B$. If we simply translate this
understanding into the language of Lewis, we see that, at least in this 
example,
we must judge as equally similar any two possible worlds which first differ
from the actual world at a given time $t$; we do not even have to inquire
whether they differ from the actual world at times later than $t$.

But why do we not have to inquire?  Let me re-phrase 
(in a perhaps slightly whimsical manner) the answer that Lewis gave
to this question in the non-relativistic context \cite{L1}. 
Consider a 
possible world $W_p$ in which a certain experiment, performed at
space-time point $r$, goes differently than in the actual world $W_a$.
Then in $W_p$ the journal article reporting the result of that
experiment is written differently than in $W_a$, and so perhaps the
experimenter receives tenure in $W_p$ but not in $W_a$, and so...
Or to extend our whimsy in a slightly different direction, we know
that any event will necessarily produce an infinite number of 
(very low energy) photons; if an event at $r$ is different in the
two worlds, the distribution of photons will be different also.
Of course, left to themselves, the photons will travel along the
(outside of the) forward light-cone of $r$; however, because of
scattering (e.\ g.\ from microwave-background photons) there will also be
effects from these photons throughout the interior of that light-cone.  
The bottom line is that, if $W_p$ differs from $W_a$ at $r$, 
and if it obeys all the laws of physics (including the arrow of time
which we are certainly not explaining here)
it will also differ (at least) at all points in $F(r)$.

This argument indicates that, for any possible world $W_p$, the set
$D_p$ will in fact coincide with the set $\overline{D_p}$.
Of course, in practice we want to specify a possible world in as
economical a way as is possible, without having to describe in detail
who receives tenure in that world, or in even more detail the effects
of an infinite number of infrared photons.  And so we {\em specify} a
possible world by saying that it differs from the actual world at a
small number of points, while we expect that it differs (at least)
in the future closure of those points. 
Lewis \cite{L1} argued that, for the non-relativistic case
(i.\ e.\ with absolute time ordering), if
$W_p$ and $W_a$ differ at time $t$, then we expect that they will
also differ at times later than $t$, but not necessarily at times
before $t$. We now see that the generalization of this conclusion
to the relativistic case is that, if they differ at space-time point $r$,
we expect them also to differ at all points in $F(r)$ (of course they
may differ at points outside $F(r)$ as well).

The definition of counterfactual I have given does allow the trivial
case in which there is no possible $\phi$-world.  However, most applications
of counterfactuals in discussions of quantum theory do assume that,
due to quantum indeterminism,
$\phi$ is in fact compatible with the past of the actual world. 
That is, in the case that $\phi$ specifies a difference from the actual
world at a single point $r$, it will usually be expected that there
will be a possible $\phi$-world which agrees with the actual world at all
points which are unambiguously earlier than $r$. This expectation
is in addition to
the expectation mentioned previously that there will
{\em not} be agreement at points which are unambiguously later than
$r$. 

\section{Some applications}
Let $\Delta$ be a region of space-time; I say that $\Delta$ is a
``$\phi$-region'' if $\Delta =\overline{D_q}$ for some $\phi$-world $W_q$.
Let $\Sigma$ be a set of $\phi$-regions; I write 
$\Sigma = \{ \Delta _{\alpha} \}$, and say that $\Sigma$ ``supports''
$\phi$ if for any $\phi$-world $W_q$ there is a 
$\Delta _{\alpha}\in \Sigma$ with $\Delta_{\alpha}\subseteq \overline{D_q}$.
(A set $\Sigma$ which supports $\phi$ is also said to be a ``coinitial''
subset of the (partially) ordered set of $\phi$-regions.\cite{Bbk})
Then if $\Sigma$ supports $\phi$: If $\Sigma$ contains only a single element
$\Delta$, then any $\phi$-world $W_p$ with $\overline{D_p}=\Delta$
is primary;
if $\Sigma$ contains only a finite
number of elements, then $\phi$ is necessarily closed;
if $W_p$ is a primary $\phi$-world, then $\overline{D_p}\in \Sigma$,
hence to evaluate the truth of a statement $\phi \; \cf \psi$ 
we need only to consider $\phi$-worlds $W_q$ with $\overline{D_q}\in \Sigma$. 
All of the examples considered in this section will have antecedents
which are supported by a set $\Sigma$ containing only a finite number of
elements; therefore these antecedents are closed, and DSTC does indeed apply.  

In many  cases in which counterfactual statements are made about
quantum systems (including the example in the first section)
the counterfactual antecedent represents a free choice.  This is
sometimes described by saying that a person can exercise free will
to choose between alternatives, but can just as well be modeled
as being controlled by the flip of a quantum coin: a device
prepares an auxiliary particle with spin aligned along the z direction,
and then measures the x-component of spin; the outcome of the
measurement then determines the ``choice''. Say that $\hat{\chi}_a$ is a
choice actually made at a space-time point $r$, and that $\hat{\chi}$
is an alternative choice that could have been made.  To say that the
choice between $\hat{\chi}_a$ and $\hat{\chi}$ was a free choice is to say
both that $\hat{\chi}$ would have been possible, and that the choice is not
correlated with anything which is space-like separated from $r$.
We can express this meaning by saying that, if $\hat{\chi}$ is a free choice,
there must exist a possible $\chi$-world $W_p$ which agrees with the
actual world everywhere outside $F(r)$; thus
$\overline{D_p}=F(r)$.

Let $\hat{\chi}$ be a free choice at $r$, and let $\chi$ be the 
proposition that the choice $\hat{\chi}$ is made;
then $\chi$ is supported by the
set consisting of the single element $F(r)$.  Therefore the primary
$\chi$-worlds are those worlds $W_p$ with $\overline{D_p}=F(r)$.
Now let $\Psi$ denote the statement ``$\chi \; \cf \psi$''
(This $\Psi$ represents the type of counterfactual statement considered
in ref.\ \cite{S}). 
Then $\Psi$ is equivalent to
the statement that $\psi$ is true in each world which agrees with the
actual world everywhere outside $F(r)$.  As a special case of this,
suppose that $\psi$ refers only to events located outside $F(r)$
(that is, events which are either unambiguously before or else are
space-like separated from $r$); then $\Psi$ is true iff $\psi$
is true in the actual world.  This result has been shown to follow
from the definition of counterfactual given in the preceding section.
Even if a $\psi$ which refers only to events located outside $F(r)$
is true in the actual world, there may well be possible
$\chi$-worlds in which $\psi$ is not true; if we had defined,
for example, the statement $\Psi$ to mean that $\psi$ was true in
{\em all} possible $\chi$-worlds, we would not have concluded
that $\Psi$ was true.

Suppose that we have several independent free choices, say $\hat{\chi}_i$
located at $r_i$, for $1\leq i \leq n$.
The proposition $(\chi_{1}\wedge \chi_{2}\wedge \cdots \wedge \chi_{n})$
is supported by the set consisting of the single element
$\bigcup_{i=1}^{n}F(r_{i})$, hence the statement
$(\chi_{1}\wedge \chi_{2}\wedge \cdots \wedge \chi_{n}) \; \cf \psi$ 
is equivalent to the statement that $\psi$ is true in every 
$(\chi_{1}\wedge \chi_{2}\wedge \cdots \wedge \chi_{n})$-world
which agrees with the actual world everywhere outside
$\bigcup_{i=1}^{n}F(r_{i})$ (that is, at all points which are not at
nor in the future of any of the $r_i$). Also, the proposition
$(\chi_{1}\vee \chi_{2}\vee \cdots \vee \chi_{n})$ is supported by the
set $\Sigma =\{ F(r_{i}\}$; if the choices are mutually space-like
separated, then any $\chi_i$-world $W_p$ with
$\overline{D_p}=F(r_{i})$ is primary, and so the
statement $(\chi_{1}\vee \chi_{2}\vee \cdots \vee \chi_{n})
 \; \cf \psi$ is equivalent to the statement that, for each $i$,
$\psi$ is true in any $\chi_i$-world which agrees with the actual
world everywhere outside $F(r_{i})$, that is, $\chi_{i} \; \cf \psi$ 
for each $i$.

The definition DSTC can also be applied to counterfactual statements
of the type discussed in ref.\ \cite{V}. Consider two 
spin-$\frac{1}{2}$ particles $A$ and $B$, and measurements performed
at a given location (so that the time-ordering is unambiguous).
Say this is the actual world: at time $t=0$ the particles are prepared
in a state of total spin zero; at time $t=1$ nothing is done; at time $t=2$
the value of $S_x$ of $B$ is measured, with result -1.  It follows,
from the fact that the total spin is zero, that the product of the values
of $S_x$ of $B$ and $S_x$ of $A$ must be -1; nevertheless, the following
statement is, on my analysis, {\em not} true: ``If $S_x$ of $A$ had
been measured at $t=1$, the result +1 would have been obtained.''
The reason this is not true is that, among worlds which already differ
from the actual world at $t=1$, a world with result +1 at $t=2$ and a
world with result -1 at $t=2$ count as equally similar to the actual world
(even though the actual result at $t=2$ was -1). However, the following
statement {\em is} true: ``If $S_x$ of $A$ had been measured at $t=1$,
and $S_x$ of $B$ measured at $t=2$ with result -1, then the result of
$S_x$ for $A$ would have been +1''. (Given the actual world as described
above, this could also be written ``If $S_x$ of $A$ had been measured at $t=1$,
and the measurement and result at $t=2$ were the same as in
the actual world, then the result of
$S_x$ for $A$ would have been +1''.)  The reason this is true is that
a world with result +1 for $B$ at $t=2$ (and hence -1 for $A$ at $t=1$)
is not a possible $\phi$-world, if $\phi$ includes the specification
that the result for $B$ was -1.    

As an example of a counterfactual statement whose antecedent 
does not involve choices, consider the case of three particles, 
each of spin $\frac{1}{2}$, located at space-time points 
$A$, $B$, and $C$. The x-component of the spin
of each is measured; denote the possible results by $a=\pm 1$, $b=\pm 1$, 
and $c=\pm 1$.  However, the particles are in an entangled state
(for example, the GHZ state \cite{GHZ}) for which the product $abc$
surely has the value -1.  Say that in the actual world, 
$a=-1,\; b=+1,\; c=+1$, and consider a statement $\Psi_{1}$, defined by
\[ \Psi_{1} :=(a=+1) \; \cf (c=+1). \] 
The proposition $(a=+1)$ is supported by the set
\[ \Sigma := \{F(A)\cup F(B), F(A)\cup F(C)\}.\]  
Denote by $W_1$ any $(a=+1)$-world
with $\overline{D_{1}}=F(A)\cup F(B)$; $W_1$ has
$a=+1,\; b=-1,\; c=+1$.  Denote by $W_2$ any $(a=+1)$-world with 
$\overline{D_{2}}=F(A)\cup F(C)$; $W_2$ has
$a=+1,\; b=+1,\; c=-1$.
If $A$, $B$, and $C$  are mutually space-like separated,
as shown in figure 2, then both $W_1$ and $W_2$ are primary, and
since $c\neq +1$ in $W_2$,
statement $\Psi_{1}$ is false.  
\begin{figure}
\epsffile{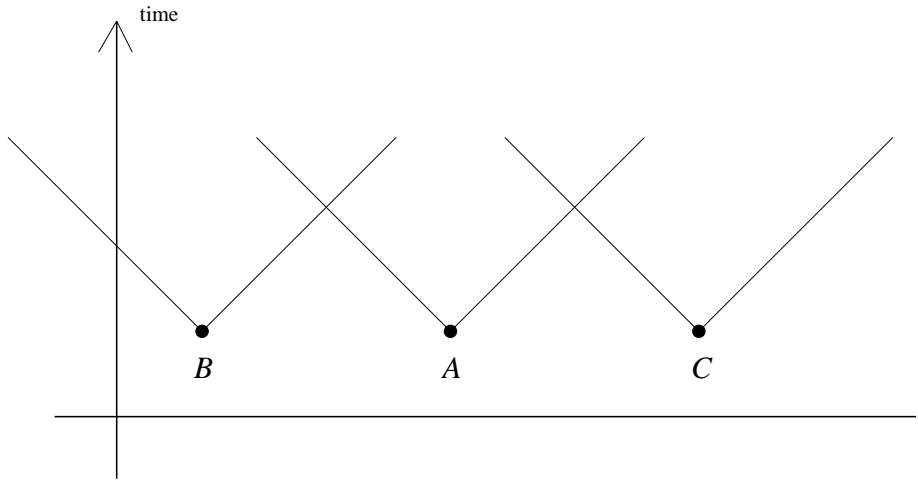}
\caption{Three space time points $A$, $B$, and $C$, mutually space-like
separated.}
\end{figure}
On the other hand, if $B$ is in the
unambiguous future of $A$, as shown back in figure 1, then 
since in this case $\overline{D_1}\subset \overline{D_2}$, 
only $W_1$ is primary, and then $\Psi_{1}$ is true.

We could also analyze this same GHZ example using the frame-dependent
definition of counterfactual mentioned in the previous section.
For that, we pick a Lorentz frame, and then say that the 
possible $(a=+1)$-world
which deviates from the actual world at the later time is more similar;
then say that the statement $\Psi_1$ defined above is true if $c=+1$ in the 
more-similar world. Take the case in which the three events are 
mutually space-like separated, as shown in figure 2.  If we pick a frame
in which the times of the events satisfy $t_{C}<t_{A}<t_{B}$
(call that frame $\alpha$), then world $W_1$ first deviates at time
$t_A$, and world $W_2$ first deviates at time $t_C$; thus 
$W_1$ is more similar to the actual world than is $W_2$, and the
statement $\Psi_1$ is true.  On the other hand, suppose we pick a frame
in which $t_{B}<t_{A}<t_C$ (frame $\beta)$; then $W_1$ first deviates
at time $t_B$, while $W_2$ first deviates at time $t_A$, and so
$\Psi_1$ is false.  Thus in this example the truth of $\Psi_1$ depends on
which frame we pick.
Finally, we can use this same GHZ example to see what would happen
if we said that a counterfactual statement is true if we can find
{\em any} Lorentz frame in which it can be ``verified'' by the above
analysis.  We would then say that $\Psi_1$ can be verified in frame
$\alpha$, and hence it is true. Also, since $c=-1$ in $W_2$,
and since in frame $\beta$ world $W_2$ is more similar than is world
$W_1$, the statement 
$(a=+1)\; \cf (c=-1)$ can be verified in frame $\beta$, and hence it
also is true.  So we wind up saying that statements  
$(a=+1)\; \cf (c=+1)$ and $(a=+1)\; \cf (c=-1)$ are both true.

In summary, this paper has proposed a definition for
space-time counterfactuals. Applications are not limited to
those in which the antecedent represents free choices, but the
implications of this definition are most simply stated for those
cases.  Some of these implications are:
if $\hat{\chi}$ represents a free choice, the counterfactual
$\chi \; \cf \psi$ is true whenever $\psi$ is true in all
possible $\chi$-worlds which agree with the actual world everywhere
outside the unambiguous future of the choice;
$(\chi_{1}\wedge \chi_{2}) \; \cf \psi$ means that $\psi$ is true in all
possible $(\chi_{1}\wedge \chi_{2})$-worlds 
which agree with the actual world everywhere outside
the unambiguous future of  either choice;
and when the two choices are space-like separated,
$(\chi_{1}\vee \chi_{2}) \; \cf \psi$ means that both 
$\chi_{1} \; \cf \psi$ and $\chi_{2} \; \cf \psi$ are true.

\section*{Acknowledgements}
I have benefitted greatly from conversations with Henry Stapp.
I would also like to acknowledge the hospitality of the
Lawrence Berkeley National Laboratory.

\end{document}